\pdfoutput=1 %arXiv admin added this line
\documentclass[copyright,creativecommons]{eptcs}
\providecommand{\event}{M. Bujorianu and M. Fisher (Eds.):\\ \hbox{\quad}Workshop on Formal Methods for Aerospace (FMA)}
\providecommand{\volume}{20}
\providecommand{\anno}{2010}
\providecommand{\firstpage}{76}
\providecommand{\eid}{8}
\usepackage{breakurl}             % Not needed if you use pdflatex only.
\usepackage{amsfonts}
  \usepackage{amsmath}
  \usepackage{amstext}
  \usepackage{amssymb}
  \usepackage{latexsym}
  \usepackage{bbm}
\usepackage{cite}
\usepackage{graphicx}
\graphicspath{{figures/}}
\usepackage{amsbsy}

\newcommand{\Rset}{\mathbb{R}}

\title{Flexible Lyapunov Functions \\and Applications to Fast Mechatronic Systems}
\author{Mircea Lazar
\institute{Dept.~of Electrical Eng., Eindhoven Univ.~of Technology,\\
       P.O.\  Box 513, 5600 MB Eindhoven, The Netherlands
}
\email{m.lazar@tue.nl}
}
\def\titlerunning{Flexible Lyapunov Functions}
\def\authorrunning{M. Lazar}
\begin{document}
\maketitle

\begin{abstract}
The property that every control system should posses is stability, which translates into safety in real-life applications.  A central tool in systems theory for synthesizing control laws that achieve stability are control Lyapunov functions (CLFs). Classically, a CLF enforces that the resulting closed-loop state trajectory is contained within a cone with a fixed, predefined shape, and which is centered at and converges to a desired converging point. However, such a requirement often proves to be overconservative, which is why most of the real-time controllers do not have a stability guarantee. Recently, a novel idea that improves the design of CLFs in terms of flexibility was proposed. The focus of this new approach is on the design of optimization problems that allow certain parameters that define a cone associated with a standard CLF to be decision variables. In this way non-monotonicity of the CLF is explicitly linked with a decision variable that can be optimized on-line. Conservativeness is significantly reduced compared to classical CLFs, which makes \emph{flexible CLFs} more suitable for stabilization of constrained discrete-time nonlinear systems and real-time control. The purpose of this overview is to highlight the potential of flexible CLFs for real-time control of fast mechatronic systems, with sampling periods below one millisecond, which are widely employed in aerospace and automotive applications.
\end{abstract}

\section{Introductory Overview}
\label{sec1}
One of the interesting problems in nonlinear control systems is the synthesis of control laws that achieve stability \cite{Sontag99, Kokotovic01}. Control Lyapunov functions (CLFs) \cite{Artstein83, Sontag83} represent a powerful tool for providing a solution to this problem. The classical approach is based on the \emph{off-line} design of an explicit feedback law that renders the derivative of the CLF negative. An alternative to this approach is to construct an optimization problem to be solved \emph{on-line}, such that any of its feasible solutions renders the derivative of a candidate CLF negative. This method can be traced back to the early results presented in \cite{Polak81}, followed by the more recent articles \cite{Primbs00, Christofides06}, where synthesis of CLFs is performed in a receding horizon fashion.

All the above works mainly deal with the continuous-time case, while conditions under which these results can be extended to sampled-data nonlinear systems using their approximate discrete-time models can be found in \cite{Grune03}. An important article on control Lyapunov functions for discrete-time systems is \cite{Kellett04SCL}. Therein, classical continuous-time results regarding existence of CLFs are reproduced for the discrete-time case. A significant relaxation in the \emph{off-line} design of CLFs for discrete-time systems was presented in \cite{Daafouz02}, where parameter dependent quadratic CLFs are introduced. Also, interesting approaches to the off-line construction of Lyapunov functions for stability analysis were recently presented in \cite{Malisoff07}, \cite{Ito08} and \cite{Ahmadi08}.

Despite the popularity of CLFs within systems theory there is still a significant gap in the application of CLFs in real-time control in general, and control of fast systems (i.e. systems with a very small sampling interval) in particular. The main reason for this is conservativeness of the sufficient conditions for Lyapunov asymptotic stability which are employed by most off-line and on-line methods for constructing CLFs.

\begin{figure}
  \centerline{\includegraphics[width=0.6\columnwidth, height=5.6cm]{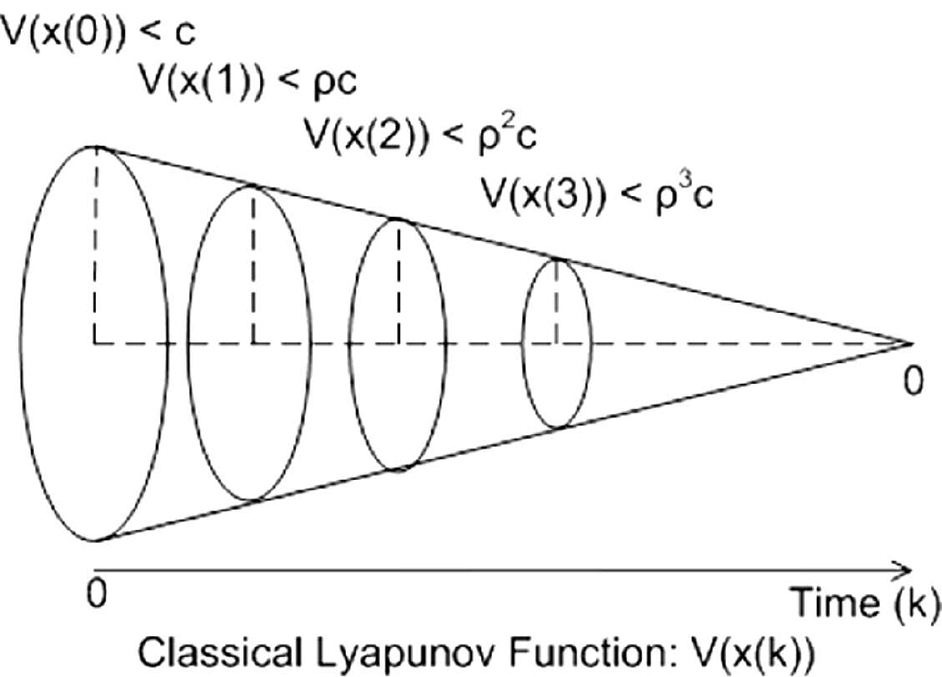}}
   \caption{A graphical illustration of classical CLFs ($\rho\in[0,1)$, $c\in\Rset_{>0}$).}
\label{fig1}
\end{figure}

To illustrate this consider the graphical depiction in Figure~\ref{fig1}.
\begin{figure}
  \centerline{\includegraphics[width=0.7\columnwidth, height=5.6cm]{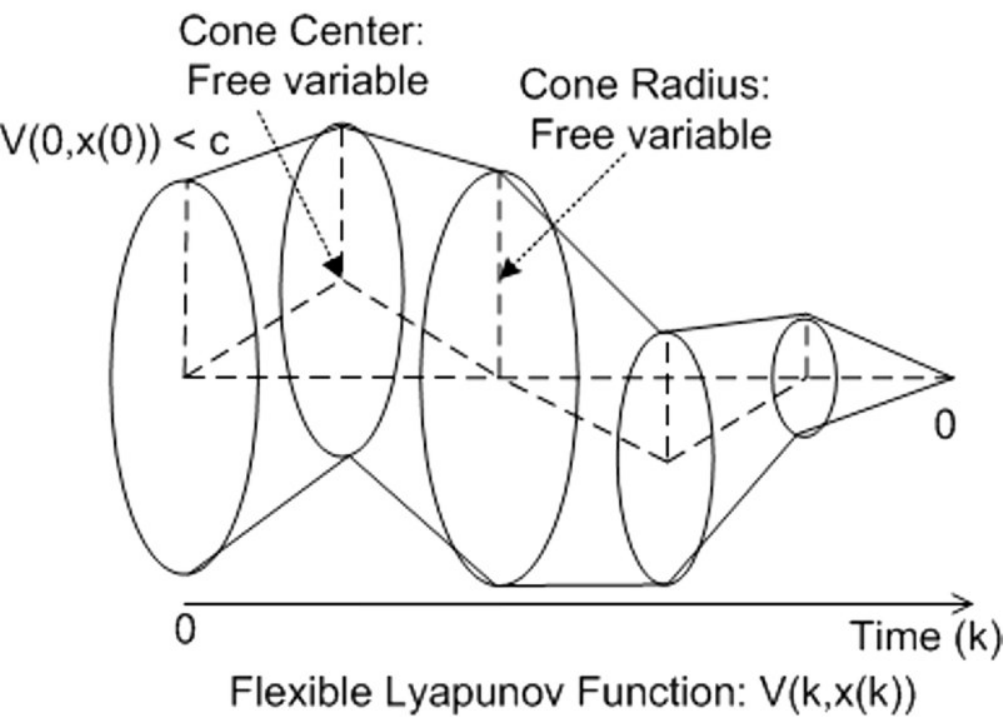}}
   \caption{A graphical illustration of flexible CLFs ($c\in\Rset_{>0}$).}
\label{fig2}
\end{figure}
Classically, a CLF enforces that the resulting closed-loop state trajectory is contained within a cone with a fixed, predefined shape, which is centered at and converges to a desired converging point. This cone is obtained by characterizing the evolution of the state via its state-space position at each discrete-time instant, with respect to a corresponding sublevel set of the Lyapunov function.
Typical examples of relevant classes of systems for which classical CLFs are overconservative are linear and nonlinear chains of integrators with bounded inputs and state constraints \cite{Mazenc97} and discontinuous nonlinear and hybrid systems \cite{Branicky}. Furthermore, in many real-life control problems classical CLFs prove to be overconservative. For example, consider the control of a simple electric circuit, such as the Buck-Boost DC-DC converter. At start-up, to drive the output voltage to the reference very fast, the inductor current must rise and stay far away (e.g., 5[A]) from its corresponding steady-state value (e.g., 0.01[A]) for quite some time. Another typical and very relevant real-life example is control of position and speed in mechatronic devices, such as electromagnetic actuators. For a given position reference, the speed must increase very fast at start-up and then return to its steady state value, which is equal to zero. In both cases enforcing a classical CLF design is obviously conservative.

Motivated by such examples, recently, in \cite{Lazar2009ACC}, a methodology that reduces the conservatism of CLF design for discrete-time nonlinear systems was proposed. Rather than searching for a global CLF (i.e. on the whole admissible state-space), therein the focus was on relaxing CLF-type conditions for a predetermined local CLF through on-line optimization problems, as it is graphically illustrated in Figure~\ref{fig2}. The goal of this overview is to highlight the potential of flexible CLFs for real-time control of fast mechatronic systems, with sampling periods below one millisecond, which are widely used in aerospace and automotive applications. This includes control of electro-magnetic actuators \cite{RalphACC09} and a real-time application to the control of DC-DC power converters.

\paragraph{Acknowledgements} This research is supported by the Veni grant ``Flexible Lyapunov Functions for Real-time Control'', grant number 10230, awarded by STW (Dutch Science Foundation) and NWO (The Netherlands Organization for Scientific Research).

\end{document}